\documentclass[twocolumn]{aastex}
\usepackage{amsmath,amssymb,amstext}
\usepackage[version = 4]{mhchem}

\usepackage{aas_macros}
\usepackage{natbib}
\usepackage{textcomp}
\usepackage{subfigure}

\begin{document}

\title{Methane in Analogs of Young Directly Imaged Exoplanets}
\author{Brittany E. Miles}
\affil{Department of Astronomy and Astrophysics, University of California, Santa Cruz}
\email{bmiles@ucsc.edu}

\author{Andrew J. Skemer}
\affil{Department of Astronomy and Astrophysics, University of California, Santa Cruz}

\author{Travis S. Barman}
\affil{Lunar and Planetary Lab, University of Arizona}

\author{Katelyn N. Allers}
\affil{Department of Physics and Astronomy, Bucknell University}

\author{Jordan M. Stone}
\affil{Steward Observatory, University of Arizona}

\begin{abstract}

We present detections of methane in R of $\sim$1300, L band spectra of VHS 1256 b and PSO 318.5, two low gravity, red, late L dwarfs that share the same colors as the HR 8799 planets. These spectra reveal shallow methane features, which indicate VHS 1256 b and PSO 318.5 have photospheres that are out of chemical equilibrium. Directly imaged exoplanets usually have redder near infrared colors than the field-age population of brown dwarfs on a color magnitude diagram. These objects along the L to T transition show reduced methane absorption and evidence of photospheric clouds. Compared to the H and K bands, L band (3 $\mu$m - 4 $\mu$m) spectroscopy provides stronger constraints on the methane abundances of brown dwarfs and directly imaged exoplanets that have similar effective temperatures as L to T transition objects. When combined with near infrared spectra, the L band extends our conventional wavelength coverage, increasing our understanding of atmospheric cloud structure. Our model comparisons show relatively strong vertical mixing and photospheric clouds can explain the molecular absorption features and continua of VHS 1256 b and PSO 318.5. We also discuss the implications of this work for future exoplanet focused instruments and observations with the \textit{James Webb Space Telescope}.

\end{abstract}

\section{Introduction}

Methane (\ce{CH4}) is fundamental to our understanding of planet formation and other processes that influence the composition of an exoplanet's atmosphere \citep{2011ApJ...743L..16O,2011ApJ...733...65B}.  Methane is detected on the majority of Solar System planets, represents a large fraction of the carbon budget in gas giant planets, and is a significant greenhouse gas produced by biological and geological processes on Earth \citep{doi:10.1146/annurev.aa.31.090193.001245,1994Icar..111..174K, 2006Natur.439..187K, 10.2307/24921496}. Detecting and constraining methane is extremely important for characterizing exoplanets as a whole.

\begin{figure}[h]
\centering
\includegraphics[width = 3.2in]{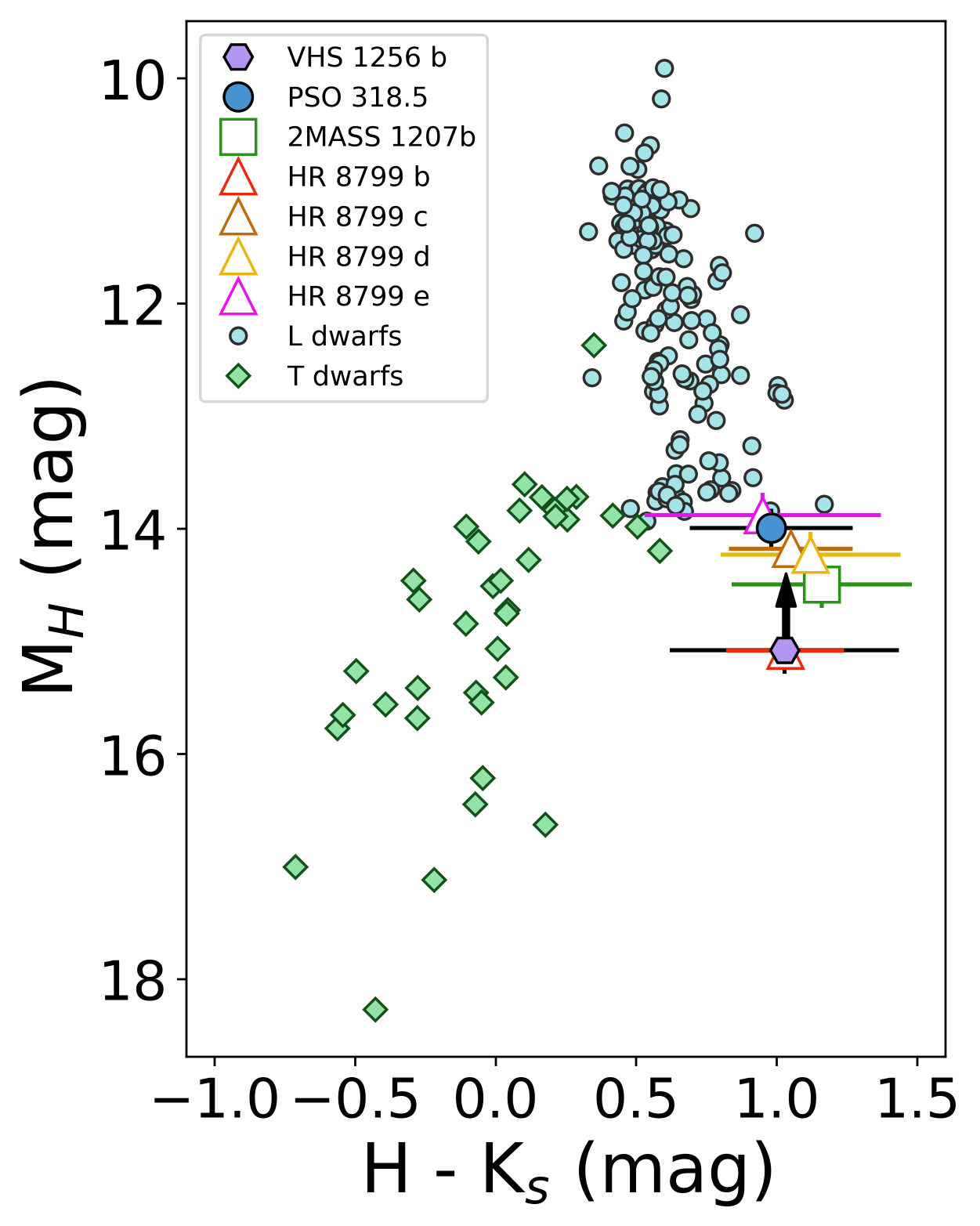}
\caption{Color magnitude diagram of L (blue circles) and T (light green diamonds) dwarfs using measured and synthesized 2MASS magnitudes compiled in \cite{2012ApJS..201...19D}. The HR 8799 planets are plotted as triangles with colored edges \citep{2008Sci...322.1348M,2010Natur.468.1080M,2009ApJ...705L.204M, 2012ApJ...753...14S}. 2MASS 1207 b, a very young, low mass companion, is plotted as a square with green edges \citep{2004A&A...425L..29C}. VHS 1256 b is plotted as a lavender hexagon and PSO 318.5 is plotted as a blue circle. The distance of VHS 1256b is not well constrained, therefore the black arrow represents a range of possible absolute magnitudes.}
\label{fig:hr-diagram}
\end{figure}

Current directly imaged gas-giants like the HR 8799 planets \citep{2008Sci...322.1348M,2010Natur.468.1080M} show much less methane in their H band spectra than is seen in the majority of similar temperature field brown dwarfs \citep{2010ApJ...723..850B,2011ApJ...733...65B}. Most directly imaged exoplanets also have red near-infrared colors compared to most field brown dwarfs, especially near the L to T transition (see Figure \ref{fig:hr-diagram}, \cite{2016ApJ...833...96L}). The L to T transition is both an evolutionary transition and a region of the near infrared color-magnitude diagram where: 1) Cloud properties rapidly alter the near infrared colors of brown dwarfs by several magnitudes over a small temperature range and 2) The dominant carbon bearing, photospheric gas transitions from carbon monoxide (\ce{CO}) to methane (\ce{CH4}) \citep{1997ApJ...491..856B,2001ApJ...556..872A, 2002ApJ...571L.151B, 2002ApJ...564..466G, 2003ApJ...585L.151T, 2004AJ....127.3553K, 2006ApJ...640.1063B, 2008ApJ...689.1327S}. The relatively red near infrared colors of directly imaged exoplanets have primarily been explained lingering photospheric clouds composed of silicate and iron condensates \citep{2011ApJ...737...34M, 2012ApJ...754..135M, 2011ApJ...729..128C}. The lack of methane is thought to be the result of disequilibrium chemistry, driven by vertical mixing interchanging cool \ce{CH4}-rich gas with warmer \ce{CO}-rich gas \citep{2011ApJ...735L..39B,2011ApJ...733...65B,2014ApJ...797...41Z}. 

To understand cloud properties and atmospheric mixing in exoplanets, broad wavelength coverage is essential for breaking degeneracies. Atmospheric models that only consider clouds can reproduce the near infrared (.8 $\mu$m - 2.3 $\mu$m) colors of exoplanets, but under-predict their fluxes from 3 $\mu$m - 4$\mu$m \citep{2010ApJ...716..417H, 2012ApJ...753...14S, 2014ApJ...792...17S, 2011ApJ...735L..39B, 2015ApJ...804...61B}. Various studies have been able to get better fits to 1 $\mu$m - 5 $\mu$m photometry and spectroscopy of 2MASSJ 1207334-393254b (2MASS 1207 b) \citep{2004A&A...425L..29C} and the HR 8799 planets by including the effects of iron and silicate clouds, patchy clouds, and non-equilibrium chemistry \citep{2008Sci...322.1348M, 2010Natur.468.1080M,2010ApJ...716..417H,2011ApJ...735L..39B, 2011ApJ...733...65B, 2015ApJ...804...61B, 2011ApJ...729..128C, 2013Sci...339.1398K,2011ApJ...732..107S, 2012ApJ...754..135M, 2012ApJ...753...14S, 2014ApJ...792...17S, 2014ApJ...795..133C, 2015ApJ...804...61B}.  

Currently, our understanding of directly-imaged exoplanet atmospheres is limited by data quality.  Directly-imaged exoplanets are difficult to observe because they are faint compared to the glare of their much brighter host stars. Young, free floating and wide separation planetary mass objects often share the same color space as directly-imaged gas giant exoplanets \citep{2013AJ....145....2F, 2016ApJ...833...96L}. These objects are significantly easier to study in detail and provide insight into the atmospheric properties of bona fide gas giant exoplanets. 

In this study, we present L band (3 $\mu$m - 4 $\mu$m) spectra of two exoplanet analogs, VHS J125601.92-125723.9b \citep[VHS 1256 b,][]{2015ApJ...804...96G} and PSO J318.5338-22.8603 \citep[PSO 318.5,][]{2013ApJ...777L..20L}.  The opacity of \ce{CH4} dominates over all other gases within the L band (peaking at $\sim$3.3 $\mu$m see \cite{2014ApJ...787...78M,2014MNRAS.440.1649Y}) allowing for an  unambiguous detection, even at low abundances.  VHS 1256b and PSO J318.5 are both low gravity objects that have similar colors to the HR 8799 planets (Figure~\ref{fig:hr-diagram}) and show no near infrared \ce{CH4} absorption. 

VHS 1256 b is a wide companion (8'' separation) to a late M dwarf binary system \citep{2016ApJ...818L..12S} making it accessible to long slit spectroscopy. The age of the VHS 1256 system is not well known, however VHS 1256 b's weak absorption from neutral gases (\ce{Na and K}) and collisionally induced absorption from hydrogen indicates a low surface gravity, which implies youth \citep{2015ApJ...804...96G}. Model fits to the optical to mid infrared photometry of VHS 1256 b suggests a very cloudy atmosphere \citep{2016ApJ...830..114R}.  PSO J318.5 is a free floating object that is a member of the $\beta$ Pictoris moving group, making it about 20 - 25 Myr old \citep{2016ApJ...819..133A}. The time series photometry and near infrared spectra of PSO 318.5 also suggest a cloudy atmosphere \citep{2015ApJ...813L..23B, 2018AJ....155...95B}.

In Sections ~\ref{sec:observations} and ~\ref{sec:data reduction} we present L band observations and reductions of VHS 1256b and PSO J318.5.  In Section  ~\ref{sec:analysis} we quantify the significance of the \ce{CH4} features and fit our L band spectra and published near-infrared spectra with atmospheric models.  We discuss the astrophysical ramifications of our results and future observations with ground-based high-contrast imagers and the \textit{James Webb Space Telescope} in Section ~\ref{sec:discussion}.  We summarize our results in Section ~\ref{sec:summary}.

\section{Observations}
\label{sec:observations}

We observed VHS 1256 b and PSO 318.5 on UT 2016 June 19 during photometric conditions using the NIRSPEC spectrograph \citep{1998SPIE.3354..566M}, an instrument on the Keck II Telescope on Maunakea. NIRSPEC was set up in low resolution mode with the KL filter and a 42\arcsec by .570\arcsec slit. Two grating positions were used to achieve a wavelength coverage from 2.9 $\mu$m to 4.4 $\mu$m with a resolution of 1300 (in low resolution mode, NIRSPEC has a nominal resolution of $\sim$ 2500 with a .380\arcsec slit). Between 4.2 $\mu$m - 4.4 $\mu$m the atmospheric transmission is zero and the KL filter only transmits out to 4.2 $\mu$m. Our observations are background limited due to the high thermal emission from the Earth's atmosphere. An integration time of 0.5 seconds with 120 co-adds was chosen for each spectral image. The spectral images were taken in an ABBA nod pattern. Telluric calibrations were taken before and after every two or three ABBA sequences for both science targets. A0V and A1V stellar spectra were taken at the same airmass as the science targets and used as telluric calibrators. The mean precipitable water vapor for our observations was 1.3 mm and the standard deviation over the night was 0.06 mm. There were only smooth changes in water vapor over each science observation block. The slit was not aligned to the parallactic angle for every observation over the course of the night, however atmospheric dispersion across the L band is negligible A summary of all NIRSPEC observations is listed in Table~\ref{tbl:obs}.

\begin{deluxetable*}{cccccccc}
\tablecaption{Summary of UT 2016 June 19 NIRSPEC Observations}
\tabletypesize{\footnotesize} 
\tablehead{
\colhead{Science} & \colhead{Grating}      & \colhead{Wavelength} & \colhead{AB}     & \colhead{Science} & \colhead{Telluric}   & \colhead{AB}      & \colhead{Telluric} \\
\colhead{Target}  & \colhead{Angle}        & \colhead{Coverage}   & \colhead{Pairs}  & \colhead{Airmass} & \colhead{Calibrator} & \colhead{Pairs}   & \colhead{Airmass} \\
\colhead{}        & \colhead{($^{\circ}$)} & \colhead{($\mu$m)}   & \colhead{(N)}    & \colhead{}        & \colhead{}           & \colhead{(N)}     & \colhead{}
}
\startdata
VHS 1256 b & 32.76 & 2.9 - 3.8 & 6  & 1.24 & HIP 63109 & 3 & 1.24  \\
VHS 1256 b & 32.76 & 2.9 - 3.8 & 4   & 1.42  & HIP 62096 & 2 & 1.42 \\
VHS 1256 b & 34.19 & 3.5 - 4.4 & 6  & 1.23 & HIP 63109 & 2 & 1.29 \\
VHS 1256 b & 34.19 & 3.5 - 4.4 & 6  & 1.39 & HIP 62096 & 2 & 1.50 
\\
\hline
PSO 318.5 & 32.76 & 2.9 - 3.8 & 6  & 1.72 & HIP 101384 & 2 & 1.72 \\
PSO 318.5 & 32.76 & 2.9 - 3.8 & 4  & 1.38 & HIP 99592  & 2 & 1.40 \\
PSO 318.5 & 32.76 & 2.9 - 3.8 & 6  & 1.35 & HIP 104810 & 2 & 1.36 \\
PSO 318.5 & 32.76 & 2.9 - 3.8 & 4  & 1.33 & HIP 109775 & 2 & 1.43 \\
PSO 318.5 & 34.19 & 3.5 - 4.4 & 6  & 1.68 & HIP 101384 & 2 & 1.64 \\
PSO 318.5 & 34.19 & 3.5 - 4.4 & 4  & 1.37 & HIP 99592  & 2 & 1.40\\
PSO 318.5 & 34.19 & 3.5 - 4.4 & 6  & 1.35 & HIP 104810 & 2 & 1.36 \\
PSO 318.5 & 34.19 & 3.5 - 4.4 & 4  & 1.32 & HIP 109775 & 2 & 1.41
\enddata

\tablecomments{The reported airmasses of the science target and calibrator observations are median values. Our observations extend out to 4.4 $\mu$m in wavelength coverage, but the transmission of the KL filter beyond 4.2 $\mu$m is zero. Each nod consists of 120 co-adds of 0.5 second exposures. }

\label{tbl:obs}

\end{deluxetable*}

\section{Data Reduction}
\label{sec:data reduction}

\subsection{Removing NIRSPEC Detector Artifacts and Background Subtraction}
Bad pixels were identified by looking at the entire night of spectral images for anomalous values (negative or extremely low/high values across sky lines) appearing at the same position.  Bad pixels in the spectral images are replaced by the average of the surrounding pixels. A-B pair nod subtraction is done to eliminate the majority of the sky lines from the traces in the calibrator and science spectral images. Each A-B pair produces a spectral image with a positive and a negative trace. The top right quadrant (redder half in low resolution mode) of NIRSPEC's detector has pattern noise that appears as a horizontal row of increased noise every eight pixels that is primarily along the spatial direction. The pattern noise response is always smaller than the sky emission and calibrator stars, but changes quickly between exposures making it difficult to remove with A-B pair nod subtraction. The width of the pattern noise was one pixel for all observations of VHS 1256 b. The first block of PSO 318.5 data in the 32.76$^{\circ}$ grating setting has one pixel wide pattern noise. The pattern noise width in the last three blocks of the same setting increased to a width of two pixels. The first half and last half of the 34.19$^{\circ}$ grating setting PSO 318.5 data have one and two pixel wide pattern noise respectively. This pattern noise was addressed in the A-B pair subtracted spectral images for each pixel by doing a linear interpolation using the pixel above and the pixel below from the affected rows.

\subsection{Spectral Image Rectification and Residual Sky Subtraction}
The spectral traces are curved in the A-B pair subtracted spectral images and are rectified before extraction. Centroids are fit along the traces of the calibrator star spectral images to estimate the deviation from a straight line and create the spatial rectification map. The spatial rectification map is applied to an A+B image of the calibrator star, where the sky lines are used to create a second rectification map for the wavelength direction. Creating rectification maps from the science targets is difficult because their traces are extremely faint. Therefore, the two spatial and spectral rectification maps calculated from the calibrator star are applied to the A-B images of the calibrator star and the A-B images of the associated science target. After nod subtracting and rectifying the calibrator and science images, there is still excess sky to be removed. At each row along the entire wavelength direction, the median of the pixels along spatial direction is subtracted off.

\subsection{Wavelength Solution}
The wavelength solution is calculated by fitting a second order polynomial to the sky emission lines along the wavelength direction of the spatially and spectrally rectified A+B calibrator images. The spectral features are identified by referencing a smoothed model of the Maunakea Sky \footnote{www.gemini.edu/sciops/telescopes-and-sites/ \\ observing-condition-constraints/ir-background-spectra} to a resolution of 1300, matching the data. For the 32.76$^{\circ}$ grating setting 18 benchmark features were used to estimate the wavelength solution. For the 34.19$^{\circ}$ grating setting, 6 benchmark features were used. We estimate the uncertainty in the wavelength solution by taking the difference between the initial wavelengths of the benchmark lines and the final wavelengths of the benchmark lines after the wavelength solution has been assigned. On average there is a .4 pixel or 2.5 * 10$^{-4}$ $\mu$m offset, but it is not significant for our analysis and is addressed by binning the spectra.

\subsection{Spectral Extraction and Error Estimation}
To find the centers and widths of the traces we take the mean of the fully rectified A-B spectral images and sum along the wavelength direction. A Gaussian is fit to the positive and negative trace profiles and the standard deviation of the profile is used to estimate the optimal radius (1.5852$\sigma$,\footnote{http://wise2.ipac.caltech.edu/staff/fmasci/GaussApRadius.pdf} typical radius was 3.5 pixels) for the boxcar extraction. To estimate the error at every wavelength element, we take the nearest 25 pixels to the left and right of the trace centroid, mask out the trace within the extraction radius, fit a first order polynomial to the counts and pixel positions, then take the variance of the residual of the fit from the row counts. The goal of this is to capture the average noise away from some baseline profile which is not necessarily constant across an entire wavelength. The variance of the residual is multiplied by the square root of the extraction width to get an error at each wavelength element.

\subsection{Telluric and Relative Flux Calibration}
The calibrators and the science targets will have their spectra influenced by the transmission of the Earth’s atmosphere, which includes absorption from gaseous methane. The absorption of Earth’s atmosphere is addressed by dividing the raw science spectrum by a telluric calibrator spectrum, which has a known response such as a black body. To account for the transmission of Earth's atmosphere, each science trace is telluric calibrated by dividing the extracted trace by an A0 or A1 star.

\begin{figure}[ht]
\centering
\includegraphics[width=3.4in]{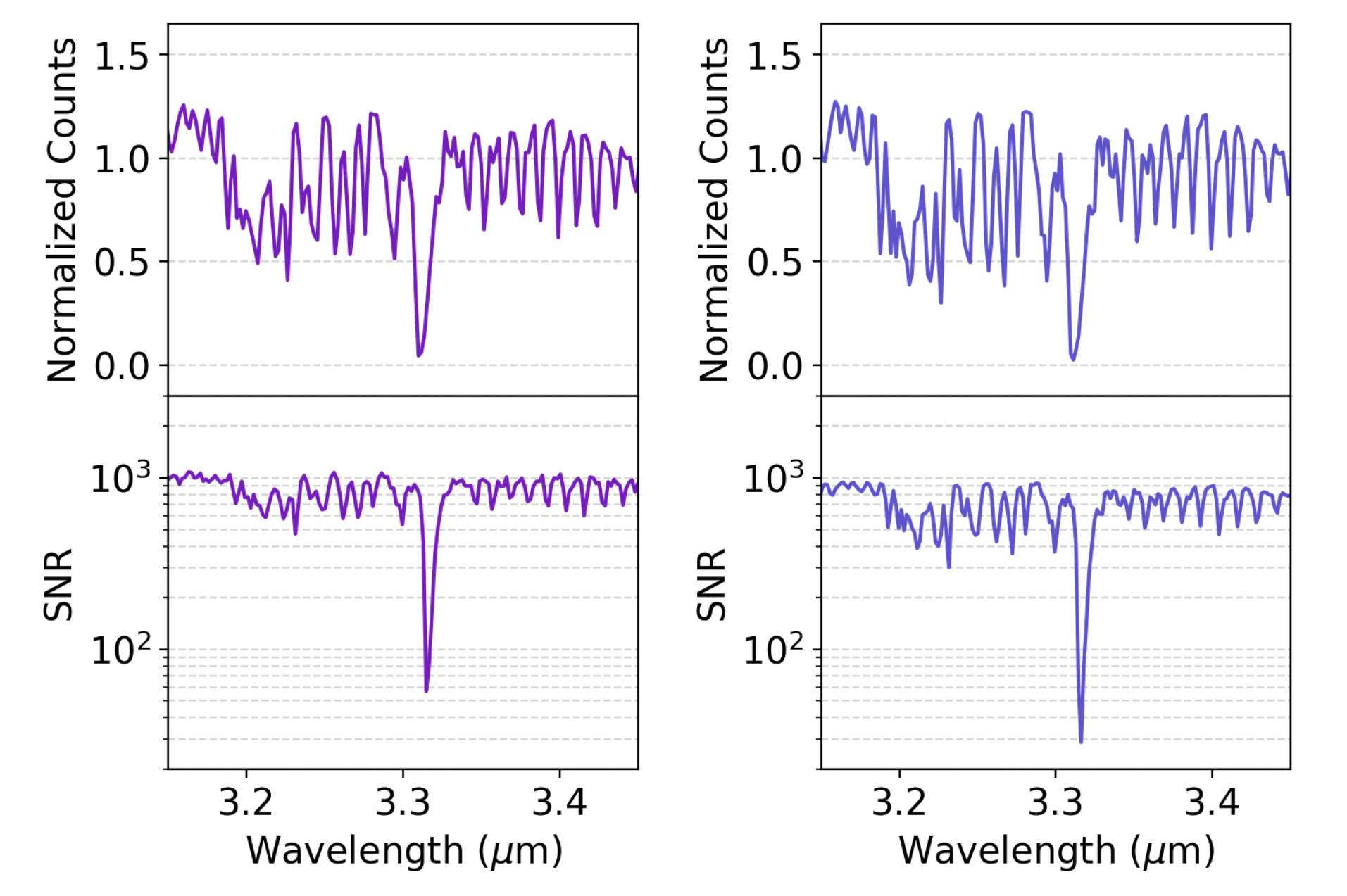}
\caption{Top: The average telluric spectra for VHS 1256 b (left) and PSO 318.5 (right), centered on the peak of the Q-branch (3.3 $\mu$m) methane feature. Bottom: The signal to noise of the average telluric spectra for VHS 1256 b (left) and PSO 318.5 (right).  At 3.31 $\mu$m, the atmospheric transmission is fairly low, however the telluric spectra have a signal to noise of 57 and 29, respectively in these troughs. }
\label{fig:telluric_spec_Q}
\end{figure}

The errors of the telluric calibrated science spectrum are calculated with standard error propagation, using error information from the extracted science trace and the telluric spectrum. The background from the science trace is the dominant source of error. The telluric calibrated line of the science target is then multiplied by Planck's Law at the appropriate temperature based on the spectral type of the calibrator star. We have two telluric calibrated spectra for every A-B image. Our science observations were bracketed by A0 and A1 stars at similar air masses for a good telluric correction, which is important especially around the methane feature at 3.3 $\mu$m. Even though the Earth's transmission is low near 3.3 $\mu$m, we still have significant signal for our calibrators (Figure ~\ref{fig:telluric_spec_Q}).

To quantify the quality of the telluric calibration, we extract the calibrator spectra taken before and after each science target block, normalize by the median, and then divide the before calibrator by the after calibrator. The ratio of the normalized ``before" calibrator to the normalized ``after" calibrator should be unity, but there are deviations, especially where the atmospheric transmission is very low. The deviation from one is interpreted as a percentage error that is included in the error of the final spectrum.

\begin{figure*}[ht]
\centering
\includegraphics[width=6.8in]{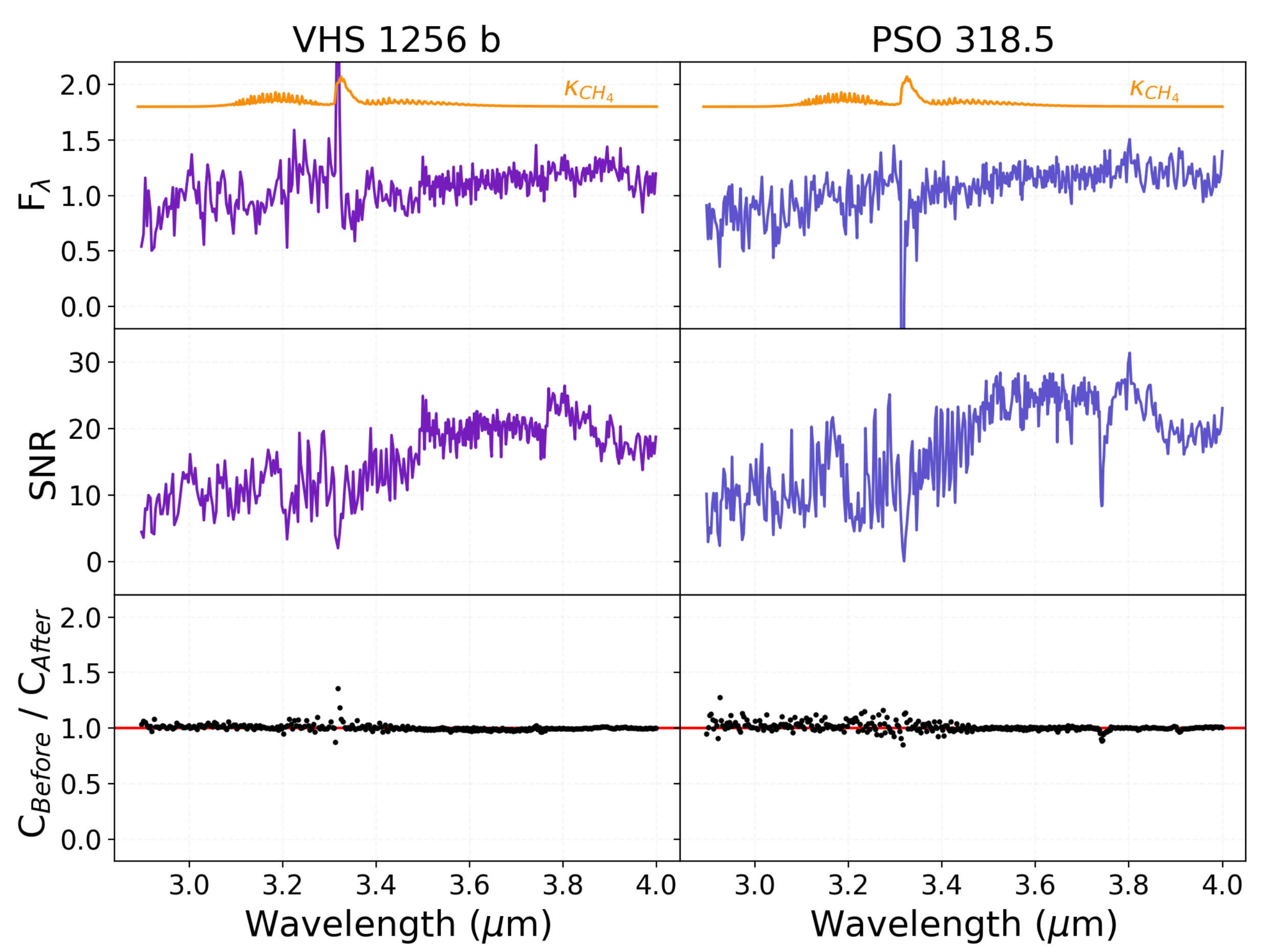}
\caption{
Top: Normalized flux (F$_{\lambda}$) plotted against wavelength for VHS 1256 b and PSO 318.5. The opacity of methane is plotted in orange above the spectra in arbitrary units on a linear scale. There is absorption from methane at $\sim$3.3 $\mu$m in both objects. Middle: The signal-to-noise vs wavelength for VHS 1256 b and PSO 318.5. Bottom: The before calibrator (C$_{Before}$) divided by the after calibrator (C$_{After}$) combined for all science observations. This is a metric of the telluric calibration quality for each object. Both objects have a standard deviation of 1\% over the entire spectrum. Across the peak of the methane feature (3.3 $\mu$m  - 3.4 $\mu$m), the standard deviation of the telluric ratio is  8\% for VHS 1256 b and 5\% for PSO 318.5. The telluric ratio deviations are included in our error propagation and reflected in the errors for the signal-to-noise plots in the middle panel.
}
\label{fig:l band spectra fig}
\end{figure*}

For each science target, the telluric and relative flux calibrated spectra from each observational block are normalized by the median value. All of the normalized data points from each observational block are combined, reordered by wavelength, and binned down to 512 pixels using a weighted average. The ratios of the ``before" and ``after" calibrators are reordered and binned down using the same weights. The final normalized spectra for VHS 1256 b and PSO 318.5 along with the telluric calibration ratios are shown in Figure~\ref{fig:l band spectra fig}.

\subsection{Absolute Flux Calibration}
\label{sec:fluxcal}

Unlike A-type stars, brown dwarfs are not well approximated by a black body due to absorption from various molecules. Therefore it is not appropriate to scale a normalized spectrum by a photometry point at the effective wavelength of a given filter. To find a better scale factor($D$) to multiply the normalized science spectrum by, we first solve for the total number of photons through a filter (and sky transmission if necessary) for a flux calibrated Vega spectrum \citep{2008AJ....135.2245R} and the normalized science spectrum.

The total number of photons($N$) measured within a filter covering a wavelength range from $\lambda_{1}$ to $\lambda_{2}$ can be approximated by
\begin{equation}
N \propto \sum_{\lambda_{1}}^{\lambda_{2}} \frac{F_{\lambda} T_{\lambda} A \tau_{exp} \lambda \Delta \lambda}  {hc}
\end{equation}
where $h$ and $c$ are the Planck constant and the speed of light respectively. The transmission through the sky and/or telescope is $T_{\lambda}$, the area of the telescope aperture is $A$, and the exposure time of the measurement is $\tau_{exp}$.

For one instrument configuration, the ratio of counts from the normalized science target and Vega should be proportional. The scale factor $D$ is then equal to 
\begin{equation}
      D = \frac{N_{sci}}{N_{Vega}} \frac{\sum_{\lambda_{1}}^{\lambda_{2}}  F_{\lambda, Vega} T_{\lambda} \lambda \Delta \lambda_{Vega}}{\sum_{\lambda_{1}}^{\lambda_{2}} F_{\lambda, sci} T_{\lambda} \lambda \Delta \lambda_{sci}}
\end{equation}
The ratio of science counts to Vega counts can be calculated using the magnitude of the science target within a given filter. To calculate the number of photons that pass through the sky (if relevant) and a filter, the transmission curves are binned down to the resolution of the spectra and then interpolated onto the same wavelengths.

The VHS 1256 b L band spectrum is flux calibrated using the Subaru Infrared Camera and Spectrograph (IRCS) L' photometry point from \cite{2016ApJ...830..114R}. PSO 318.5 has a  Wide-field Infrared Survey Explorer (WISE) \citep{2010AJ....140.1868W} W1 photometry point, but the W1 filter transmits light blueward of the L band spectra. Flux calibrating the PSO 318.5 L band spectrum is not possible without making assumptions about the continuum shape, therefore the scale factor ($D$) is allowed to be offset by up to 10\% while model fitting.

\section{Analysis}
\label{sec:analysis}

\subsection{Detection and Significance of Methane in L band Spectra}
\label{sec:L band spectra}

\begin{figure}[ht]
\centering
\includegraphics[width = 3.1in]{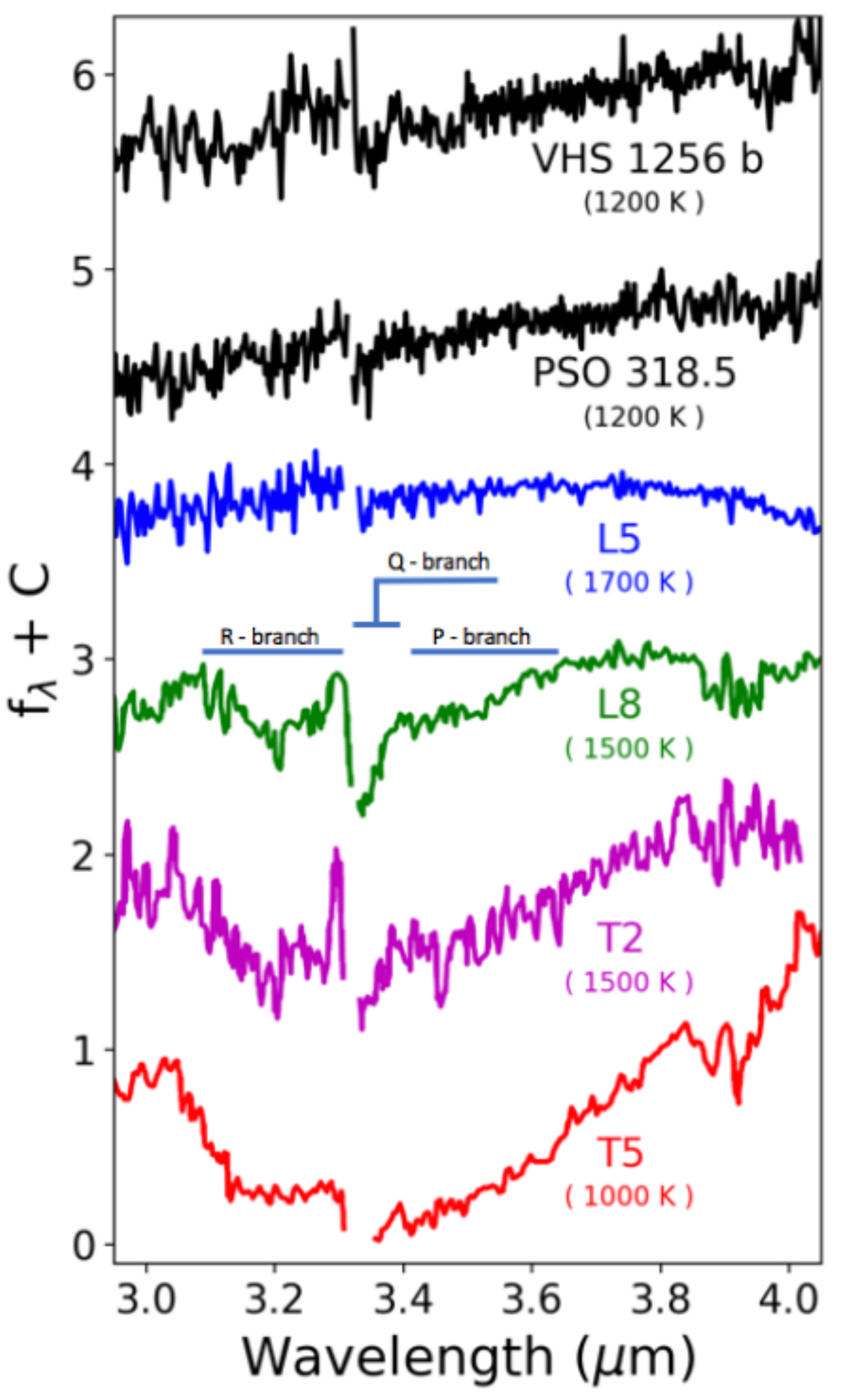}
\caption{
Normalized spectra of VHS 1256 b and PSO 318.5 (with SNR $<$ 3 data points removed) along with L band spectra of 2MASS J1507-1627 (L5), DENIS J0255-4700 (L8), SDSS J1254-0122 (T2), and 2MASS J0559-1404 (T5) from \cite{2005ApJ...623.1115C}. All of the spectral type identifications were done in the near infrared. Each spectrum is normalized at 3.8 $\mu m$ and offset by an arbitrary constant. The best fit temperature estimate for VHS 1256 b is from \cite{2016ApJ...830..114R}, for PSO 318.5 the estimate is from \cite{2013ApJ...777L..20L}. The temperatures of the brown dwarfs from \cite{2005ApJ...623.1115C} are estimated using the temperature vs. spectral type relationship derived in \cite{2004AJ....127.3516G}. The P, Q, and R branch portions of methane absorption are labeled on the L8 brown dwarf. VHS 1256 b and PSO 318.5 have weak Q-branch methane features and show less methane absorption than brown dwarfs of similar temperatures.}
\label{fig:templates}
\end{figure}

Methane is predicted to be the only significant absorber across the L band for $\sim$1300K objects \citep{2014ApJ...787...78M}. Methane absorption becomes deeper and broader through the L to T transition, especially around the Q - branch band head of methane that peaks in opacity at $\sim$ 3.3 $\mu$m \citep{2000ApJ...541L..75N, 2005ApJ...623.1115C, doi:10.1146/annurev.astro.42.053102.134017}. VHS 1256 b and PSO 318.5 have features around $\sim$3.3 $\mu$m that resemble absorption from methane. Despite their cool photospheric temperatures, these features appear shallow relative to other late L dwarfs and early T dwarfs (Figure \ref{fig:templates}). Unlike field L dwarfs, the spectral slopes of VHS 1256 b and PSO 318.5 show increasing flux density at longer wavelengths along the L band. These slopes resemble the spectral shapes of the T dwarfs, but with less methane absorption.

\begin{figure}[ht]
\centering
\includegraphics[width = 3in]{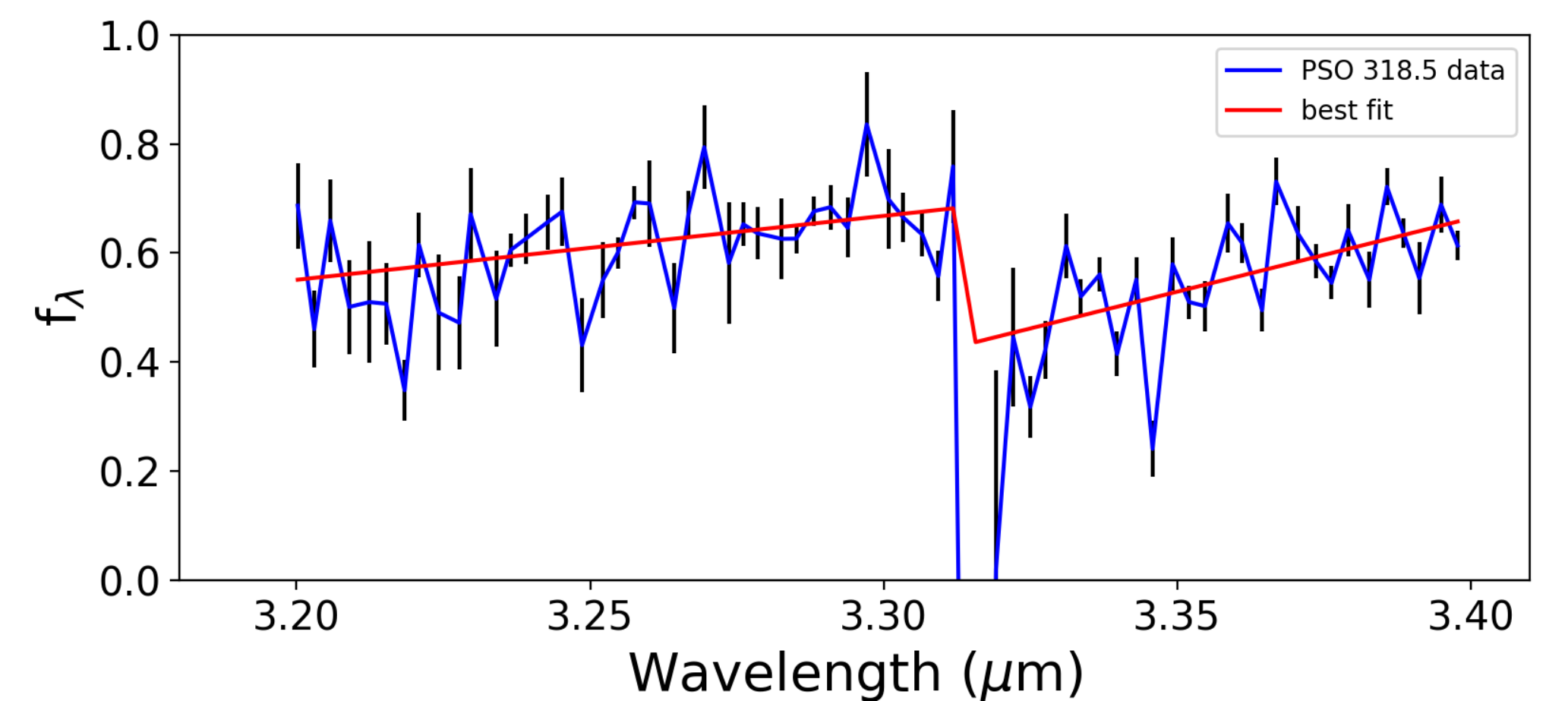}
\caption{To quantify the significance of the methane absorption feature at $\sim$ 3.3 $\mu$m, we fit a one-line model and a two-line model with a characteristic break at $\lambda_{0}$ to the data of VHS 1256 b and PSO 318.5. The two-line model produces better fits to our data, showing that the absorption is statistically significant. In this figure, we show the best fit two-line model (red) to a subset of the PSO 318.5 spectrum (blue).}
\label{fig:bestfit}
\end{figure}

We provide evidence for the methane detection by showing that a best fit line to the data from 3.2 $\mu$m - 3.4 $\mu$m produces a worse fit than a simple two-line model with a break at the characteristic wavelength ($\lambda_{0}$) over the same wavelength range. A two-line model does not accurately represent the complex absorption feature of methane, but we want to rule out a smooth continuum before comparing to models with varying methane abundances later in Section~\ref{sec:l inst}. Methods that quantify absorption, such as the equivalent width or spectral indices require knowledge of the continuum and that is difficult to account for across the L to T transition as methane absorption gets broader. The 3.2 $\mu$m - 3.4 $\mu$m range was chosen for fitting, because that is where methane absorption appears most consistent from the mid-L to mid-T dwarf range (Figure \ref{fig:templates}). The two-line model $y(\lambda$) is described by the function
\begin{equation}
y(\lambda) =  \left\{
\begin{array}{lll}
      m_{1}\lambda + b_{1}  & for & \lambda \leq \lambda_{0} \\
      m_{2}\lambda + b_{2}  & for & \lambda > \lambda_{0} \\
\end{array} 
\right\}
\end{equation}
and the coefficients are fit using the  Markov Chain Monte Carlo (MCMC) method with the python package \texttt{emcee} \citep{2013PASP..125..306F}. An example of the two-line model can be seen in the best fit for PSO 318.5 in Figure~\ref{fig:bestfit}.

The relative model quality of the one and two-line best fits were assessed by a $\Delta \chi^{2}$ value and the Akaike Information Criterion (AIC). The difference between the one and two-line best fit $\chi^{2}$ values produce a  $\Delta \chi^{2}$ value of 74.6 and 100.3 for VHS 1256 b and PSO 318.5 respectively. For VHS 1256 b, the best fit two-line model has an AIC of 159.0 and the one-line model has an AIC of 227.7. For PSO 318.5, the best fit two-line model has an AIC of 181.4 and the one-line model has an AIC of 275.7. For both objects, the two-line model produces a lower AIC and a better quality fit to the data. 

\begin{figure}[ht]
\centering
\includegraphics[width = 3.1in]{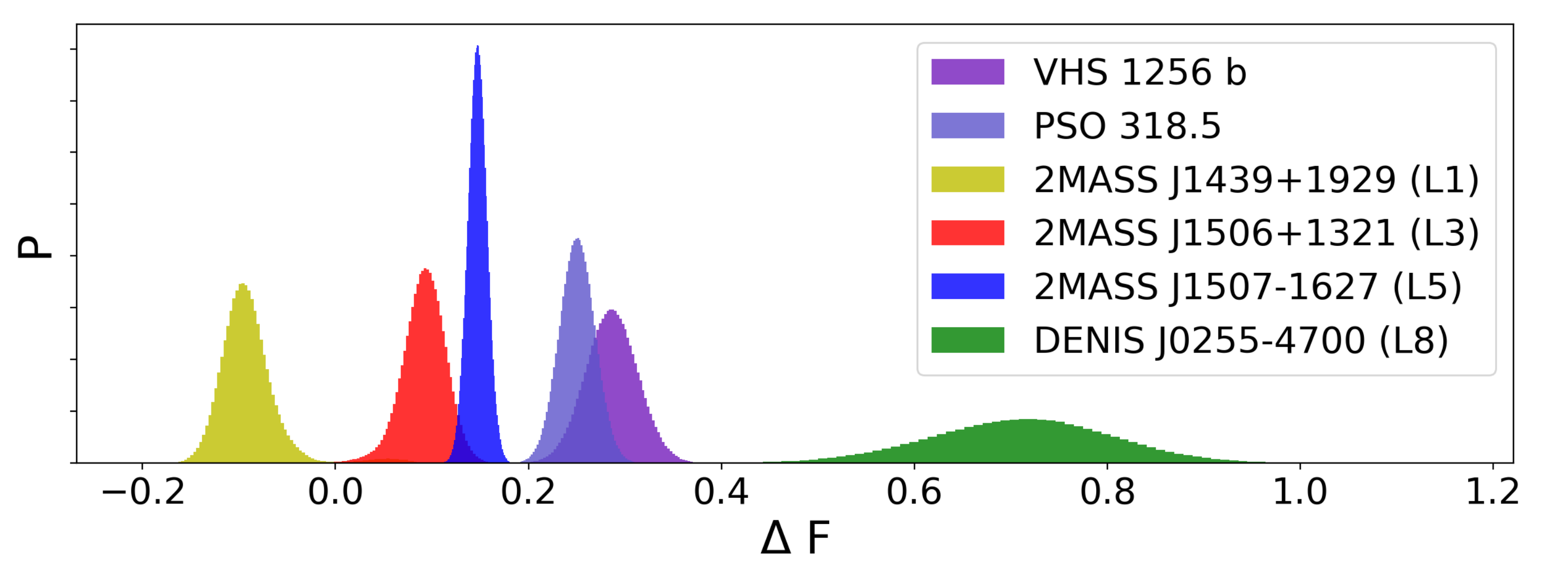}
\caption{Probability distributions of the depth parameter ($\Delta$F) for (left to right)2MASS J1439+1929, 2MASS J1506+1321,  2MASS J1507-1627, PSO 318.5, VHS 1256 b, and DENIS J0255-4700. Positive $\Delta$F values correspond to methane absorption. The mean values for each distribution (left to right) are -0.02, 0.09, 0.15, 0.25, 0.34, and 0.66. VHS 1256 b and PSO 318.5 have methane absorption distributions that are intermediate between the distributions of 2MASS J1507-1627 (1700K) and DENIS J0255-4700 (1500K). }
\label{fig:delF}
\end{figure}

The depth parameter ($\Delta$F) is a derived quantity from the two-line model that is the difference between $y_{1}(\lambda_{0})$ and $y_{2}(\lambda_{0})$.  A higher value of  $\Delta$F corresponds to a deeper absorption feature. The posterior distributions of the depth parameter for VHS 1256 b and PSO 318.5 are compared with two L dwarfs, 2MASS J1507-1627 and DENIS J0255-4700, featured in Figure~\ref{fig:delF}. Two other brown dwarfs, 2MASS J1439+1929 (L1) and 2MASS J1506+1321 (L3) are also plotted for reference. The mean values of $\Delta$F for VHS 1256 b and PSO 318.5 are inconsistent with a straight line ($\Delta$F = 0.0) by 10.5$\sigma$ and 13.3$\sigma$ respectively. Nearly the entire probability distribution of VHS 1256 b and PSO 318.5's $\Delta$F parameter lies in between the distributions of the 1700K (L5) brown dwarf and the 1500K (L8) brown dwarf. This quantitatively shows that VHS 1256 b and PSO 318.5 have methane, but show less methane absorption than hotter brown dwarfs. There is significant overlap between the $\Delta$F distributions of VHS 1256 b and PSO 318.5, but the mean $\Delta$F of VHS 1256 b is slightly higher.

The depleted methane features of VHS 1256 b and PSO 318.5 are evidence for non-equilibrium chemistry. Previous work has shown that vertical mixing can displace methane from the photosphere with warmer, carbon monoxide dominated gas \citep{2007ApJ...669.1248H, 2014ApJ...797...41Z}. In Section ~\ref{sec:models}, we further quantify VHS 1256 b and PSO 318.5's methane absorption features and estimate the magnitude of their vertical mixing using custom atmospheric models.

\subsection{Published Optical/ Near Infrared Studies of VHS 1256b and PSO 318.5}

VHS 1256 b and PSO 318.5 both have optical to mid-infrared photometric measurements and  near infrared spectra that have been published. For our analysis, we applied the absolute flux calibration outlined in Section \ref{sec:fluxcal} to the normalized near infrared spectra of VHS 1256 b and PSO 318.5 from \cite{2015ApJ...804...96G} and \cite{2013ApJ...777L..20L}. The normalized optical/near infrared spectrum from \cite{2015ApJ...804...96G} is broken up into optical/J, H, and K portions that are flux calibrated separately using the J, H, and K VISTA InfraRed CAMera (VIRCAM) photometry. Two near-infrared spectra of PSO 318.5 were presented in \cite{2013ApJ...777L..20L}, but for this work we do our analysis with the Gemini Near-Infrared Spectrograph (GNIRS) data because it has a higher spectral resolution.  The near-infrared spectrum of PSO 318.5 was flux calibrated using the K band MKO photometry \citep{2013ApJ...777L..20L} measured using WFCAM on the UK Infrared Telescope. 

Both objects have Two Micron All-Sky Survey (2MASS) measurements, but the 2MASS photometry is not consistent with the VISTA (VHS 1256 b) or MKO (PSO 318.5) photometry. The VISTA and MKO photometry are used for flux calibration because they produce less scatter in the final spectra when scaling by the different infrared photometry points (J, H, K) and are more precise than the 2MASS photometry. Photometric variability studies have been done on PSO 318.5 and similar objects \citep{2017AstRv..13....1B} and this effect will be discussed later in Section~\ref{sec:jwst} although we note that the measured variability cannot explain the photometric discrepancy discussed here.

\subsection{Description of Models and Fitting}
\label{sec:models}

The models used to fit the data were calculated using a method similar to the one outlined in \cite{2011ApJ...735L..39B} with the \texttt{PHOENIX} atmospheric code and updated \ce{CH4}, \ce{CO2}, and \ce{NH3} line lists. The intermediate cloud model (ICM) has a base pressure set by equilibrium chemistry and the vertical extent of the cloud is parameterized by a pressure value, above which the number density of cloud particles falls off exponentially. The cloud particle sizes follow log-normal distribution characterized by a median grain size. The composition of the cloud is dependent on equilibrium chemistry, resulting in clouds composed primarily of silicates (Ex. Enstatite (\ce{MgSiO3}), Forsterite (\ce{MgSiO4})).

A grid similar to \cite{2011ApJ...735L..39B} was used to fit the parameters of VHS 1256 b and PSO 318.5 to confirm that the inferred temperatures and gravities were similar to previous studies on these two objects. Vertical mixing is parameterized by the vertical eddy diffusion coefficient $K_{zz}$ which is defined as
\begin{equation}
    K_{zz} = \frac{L^{2}}{\tau_{dyn}}
\end{equation}
where $\tau_{dyn}$ is the dynamical time scale and $L$ is the length scale of diffusion. The length scale in this work is approximately a third of the pressure scale height as outlined in \cite{1998Icar..132..176S}.

Another model grid with slightly finer parameter sampling was created spanning an effective temperature of 1000 K to 2000 K (every 50 K), log(g) from 3.0 - 4.0 (every 0.25 decs),  mean grain sizes of (0.5, 1, 2, 3 and 5 $\mu$m) and 5 different cloud heights. The abundances of these models are assumed to be solar according to \cite{2009ARA&A..47..481A}. After an initial best fit was found using $\chi^{2}$ values, the cloud distribution, effective temperature and \ce{CH4} mixing ratio above the quenching point was tuned for a better fit.

\begin{figure}[h]
\centering
\includegraphics[width = 3.3in]{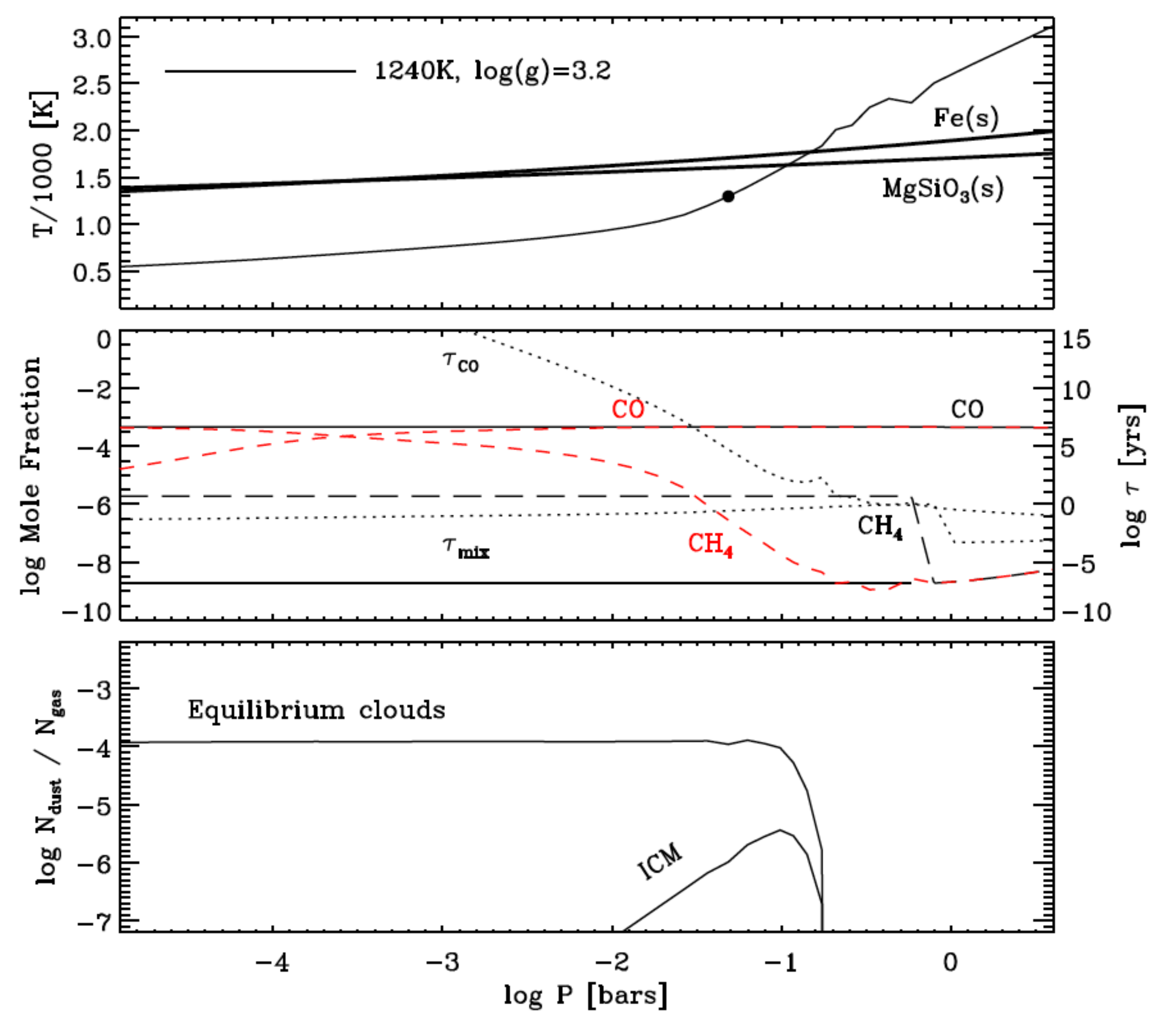}
\caption{Atmospheric properties for VHS 1256 b and PSO 318.5. Top: Temperature versus pressure profile. The black dot approximates the location of the photosphere and the condensation curves of Enstatite (MgSiO$_{3}$) and Iron (Fe) are represented as  thick black lines. Middle: \ce{CO} and \ce{CH4} mole fractions for equilibrium
(dashed, red), non-equilibrium (solid, black, with $K_{zz}$ of $10^{8}$ cm$^{2}$ s$^{-1}$) chemistry, and best fit non-equilibrium with enhanced \ce{CH4} (dashed, black, with $K_{zz}$ of $10^{8}$ cm$^{2}$ s$^{-1}$). Chemical and mixing timescales are also plotted (dotted lines). Bottom: dust to gas ratio for the intermediate cloud model (ICM), used for the best fit and the pure equilibrium cloud model.
}
\label{fig:clouds - mixing}
\end{figure}

\begin{figure*}[ht]
\centering
\includegraphics[width=7in]{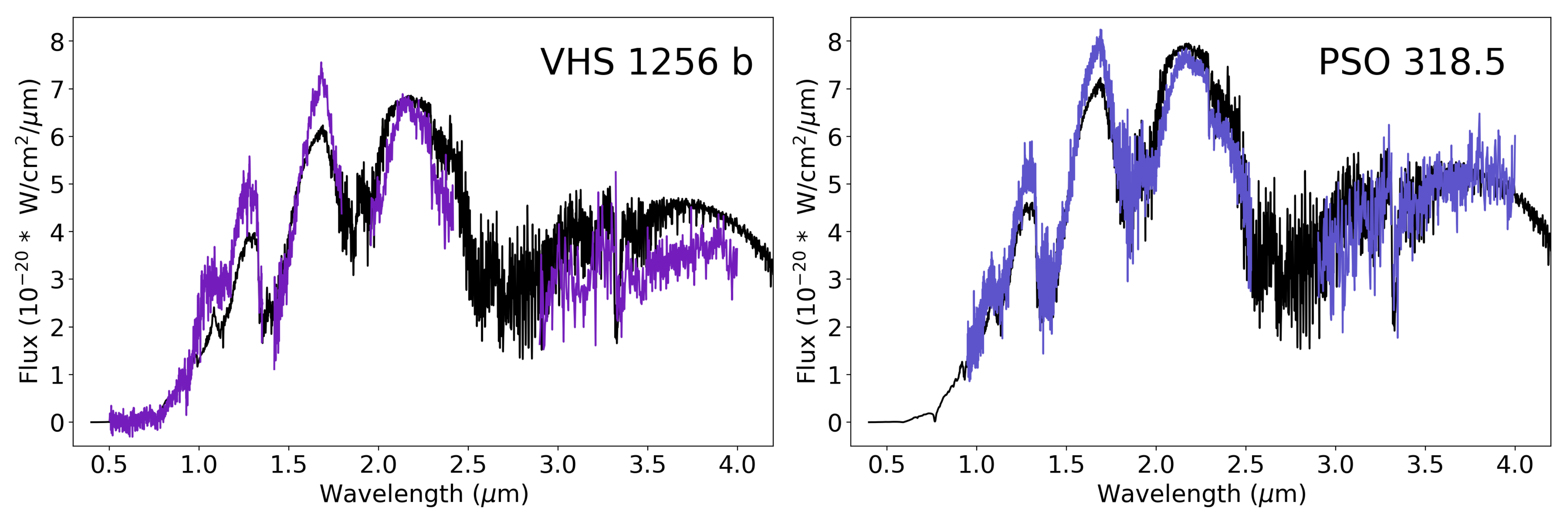}
\caption{Near infrared and L band spectra for VHS 1256 b and PSO 318.5. The near infrared spectra for VHS 1256 b  and PSO 318.5 are from \cite{2015ApJ...804...96G} and \cite{2013ApJ...777L..20L}. The best fit, low gravity (log(g) = 3.2), 1240 K model is shown in black in both panels.}
\label{fig:nir and l spectra}
\end{figure*}

The best fit model for PSO 318.5 also applies to VHS 1256 b because the objects share similar spectral features. The best fit model has its structure shown in Figure~\ref{fig:clouds - mixing} and the spectrum is shown along with the data in Figure~\ref{fig:nir and l spectra}. VHS 1256 b and PSO 318.5 each have a best fit temperature of 1240 K and a best fit $K_{zz}$ of $10^{8}$ cm$^{2}$ s$^{-1}$. Above the quenching point, the mixing ratio of \ce{CH4} was increased by two orders of magnitude to $2.4 * 10^{-6}$ for PSO 318.5 and $3.5 * 10^{-6}$ for VHS 1256 b to get a better fit. From about 1 to .1 bar, the abundance of \ce{CH4} is two orders of magnitude higher in the best fit non-equilibrium case than the equilibrium case. Above .1 bar, the best fit non-equilibrium model has one to two orders of magnitude less methane than the equilibrium case (Figure~\ref{fig:nir and l spectra}, Middle Panel). The photospheres have clouds with a median grain size of 0.5 $\mu$m. An increased \ce{CH4} mixing ratio also implies that a model with a smaller $K_{zz}$ value could also explain the spectra. 

With an assumed distance of 22 pc \citep{2016ApJ...833...96L}, the inferred radius of PSO 318.5 is 1.3 $R_{Jup}$, which is consistent with PSO 318.5 being a young planetary mass object \citep{1998A&A...337..403B}. With an assumed distance of 17 pc \citep{2016ApJ...818L..12S}, the implied radius of VHS 1256 b is 0.9 $R_{Jup}$, which is inconsistent with youth \citep{1998A&A...337..403B}. The parallax and photometric distance estimates of the binary host are discrepant\citep{2016ApJ...818L..12S}, therefore there is large uncertainty on the actual radius of VHS 1256 b. VHS 1256 b has very similar spectral features as PSO 318.5 and to get a similar radius, VHS 1256 b could possibly be as far as 20 parsecs. If VHS 1256 b is actually at 20 parsecs, its place on the color magnitude diagram would move up closer to the absolute magnitude of PSO 318.5. Constraints on the distance and age of VHS 1256 b will allow for evolutionary constraints and a comparative analysis on the atmospheric properties and evolution of red, dusty objects. A summary of our best fit and previously published parameters for VHS 1256 b and PSO 318.5 are shown in Table~\ref{tbl:params}, along with parameters for similar objects.

\begin{deluxetable}{cccccc}
\tablecaption{Atmospheric Parameters of Low Gravity Objects}
\tabletypesize{\footnotesize} 
\tablehead{
  \colhead{Object} & \colhead{T$_{eff}$}   &  \colhead{log(g)} & \colhead{Radius} & \colhead{K$_{zz}$} &\colhead{Ref} \\
  \colhead{} & \colhead{(K)}   &  \colhead{} & \colhead{ (R$_{Jup}$) } & \colhead{ (cm$^{2}$ s$^{-1}$) } &\colhead{} 
}
\startdata
\textbf{VHS 1256 b} & \textbf{1240} & \textbf{3.2} & \textbf{0.9} & \textbf{10$^{8}$} & \textbf{1}\\
VHS 1256 b & 800 - 1000 & 3.5 - 4.5 & 1.34 - 1.8 & N/A & 2,3\\
\hline
\textbf{PSO 318.5} & \textbf{1240} & \textbf{3.2} & \textbf{1.3} & \textbf{10$^{8}$} & \textbf{1}\\
PSO 318.5 & 1210 & 4.21 & 1.40 & N/A & 2\\
\hline
HR 8799 c & 1100 & 3.5 & 1.40 & 10$^{8}$ & 5\\
HR 8799 d & 1100 & 3.5 & 1.40 & 10$^{8}$ & 5 \\
2M1207 b  & 1000 & 4.0 & 1.5  & 10$^{8}$ & 6
\enddata
\tablecomments{The atmospheric properties of VHS 1256 b, PSO 318.5, and other low gravity brown dwarfs and exoplanets for comparison. The adopted parameters of this work are bolded. The distance of VHS 1256 b is not well constrained therefore a range of parameters are presented for previously published work. The parameters for HR 8799 c and d are from the grid that gave the best fit for data across the 3$\mu$m - 5$\mu$m range in \cite{2018AJ....155..226G}. References: 1 - this work, 2 - \cite{2013ApJ...777L..20L}, 3 - \cite{2016ApJ...830..114R}, 4 - \cite{2015ApJ...804...96G}, 5 - \cite{2018AJ....155..226G}, 6 - \cite{2011ApJ...735L..39B}.  }
\label{tbl:params}
\end{deluxetable}

\section{Discussion}
\label{sec:discussion}

\subsection{Strong Vertical Mixing in VHS 1256 b and PSO 318.5}
In Section~\ref{sec:L band spectra} we showed that VHS 1256 b and PSO 318.5 have less L band methane absorption than similar temperature brown dwarfs. This is evidence that their photospheres are not in chemical equilibrium. Strong vertical mixing can lower the amount of methane visible in the infrared spectra of low-gravity brown dwarfs and exoplanets \citep{2007ApJ...669.1248H,2011ApJ...735L..39B,2015ApJ...804...61B}. Methane's cross section per volume peaks at 3.3 $\mu$m, therefore L band spectra provide excellent constraints on methane abundances and vertical mixing. 

A relatively high eddy diffusion coefficient was needed in our models ($K_{zz} = 10^{8}$ cm$^{2}$ s$^{-1}$) to match the depths of the 3.3 $\mu$m methane features. Assuming mixing length theory and that all of the convective flux from heat transport is contributing to mixing, the upper limit on $K_{zz}$ is expected to range from $10^{8}$ cm$^{2}$ s$^{-1}$ -  $10^{9}$ cm$^{2}$ s$^{-1}$ for low gravity, self luminous, gas giant planets \citep{2014ApJ...797...41Z}.

\subsection{Comparison with 3 \texorpdfstring{$\mu$m}{} - 4 \texorpdfstring{$\mu$m}{} photometry of HR 8799 c and d}

\begin{figure}[h]
\centering
\includegraphics[width = 3in]{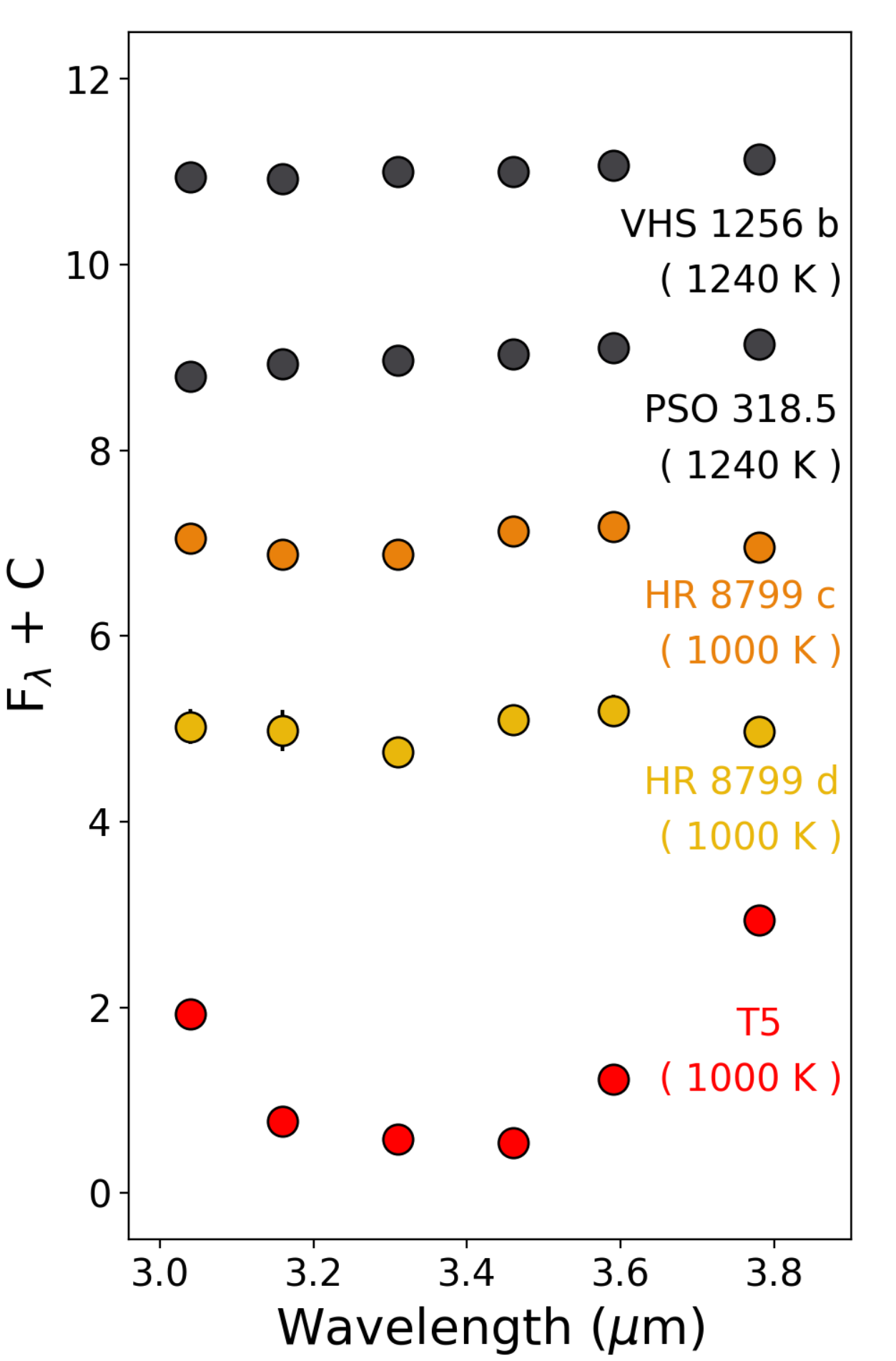}
\caption{The top data points(black) are normalized 3 $\mu$m - 4 $\mu$m photometry of VHS 1256 b and PSO 318.5 calculated using the narrow band filter profiles from \cite{2014ApJ...792...17S}. The temperatures of VHS 1256 b and PSO 318.5 are now derived from our model fits. The narrow band filter photometry for HR 8799 c and d are from \cite{2014ApJ...792...17S} and are plotted as orange and yellow points. The narrow band photometry of 2MASS J0559-1404 (featured in Figure~\ref{fig:templates}, from \cite{2005ApJ...623.1115C}) is plotted in red. Relative to VHS 1256 b and PSO 318.5, HR 8799 c and d show deeper absorption at the 3.31 $\mu$m photometry point, which suggests that they may have stronger methane absorption. However, the absorption is still more shallow than seen in the similar temperature brown dwarf 2MASS J0559-1404, indicating that the HR 8799 c and d atmospheres are out of chemical equilibrium.} 
\label{fig:hr8799compare}
\end{figure}

As color-magnitude analogs of HR 8799 c and d, the spectra of VHS 1256 b and PSO 318.5 provide insight into the methane abundances and vertical mixing of these exoplanetary atmospheres. Using L/M band mid-InfraRed Camera (LMIRcam) on the Large Binocular Telescope, \cite{2014ApJ...792...17S} took adaptive optics corrected images of the HR 8799 system in six different custom narrow band filters. The images resulted in 3 $\mu$m to 4 $\mu$m photometry of HR 8799 c and d. The NIRSPEC L band spectra of VHS 1256 b and PSO 318.5 are convolved with the filters used in \cite{2014ApJ...792...17S} to get synthetic photometry for comparison with HR 8799 c and d. In Figure~\ref{fig:hr8799compare}, we show the normalized photometry of VHS 1256 b, PSO 318.5, HR 8799 c and d, along with 2MASS J0559-1404, a $\sim$1000 K, T5 brown dwarf. The photometry for VHS 1256 b, PSO 318.5, HR 8799 c and d are relatively flat, but there are slight differences. The spectral slopes of HR 8799 c and d  start to decrease in flux past the 3.59 $\mu$m photometry point, whereas the spectral slopes for VHS 1256 b and PSO 318.5 increase across the entire L band. This may be due to slightly different cloud or temperature-pressure properties between these objects. Across 3.31 $\mu$m, the spectral slopes of VHS 1256 b and PSO 318.5 appear as straight lines due to their weak methane signatures. HR 8799 c and d show a slight depression at the 3.31 $\mu$m photometry point, which indicates that they may have more photospheric methane and weaker vertical mixing than VHS 1256 b and PSO 318.5. VHS 1256 b, PSO 318.5 and HR 8799 c and d all have less methane absorption relative to 2MASS J0559-1404 despite having similar photospheric temperatures, which indicates that all four objects are out of chemical equilibrium.  In previous work by \cite{2010ApJ...716..417H} and \cite{2014ApJ...792...17S}, non-equilibrium chemistry and photospheric clouds were needed to fit the near infrared and mid infrared photometry of the HR 8799 c and d and our model analysis on analogs VHS 1256 b and PSO 318.5 is consistent with this view.

\subsection{Using L Band spectroscopy for Detecting Methane in Exoplanets and Brown Dwarfs}
\label{sec:l inst}

\begin{figure}[ht]
\centering
\includegraphics[width = 3.2in]{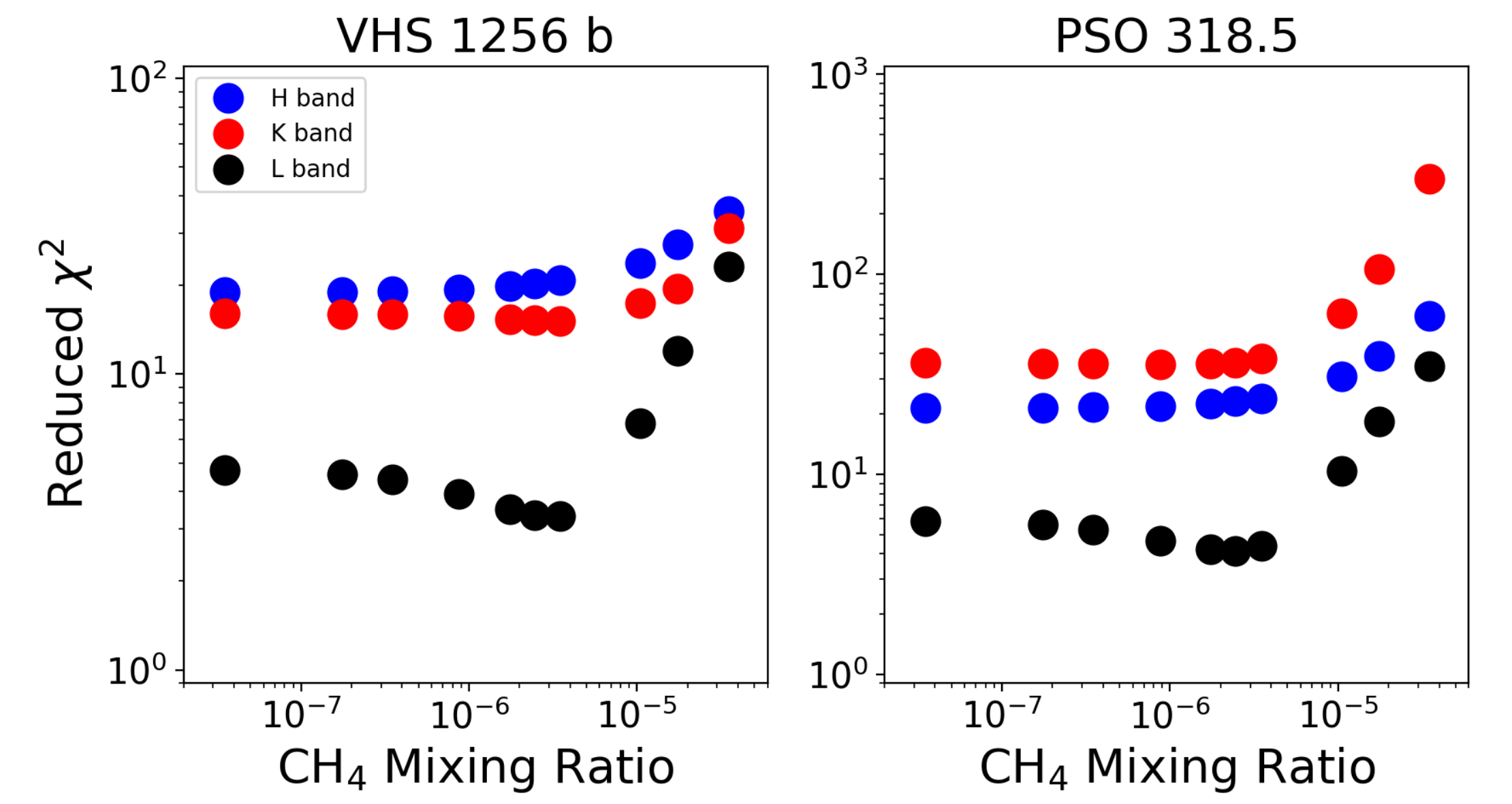}
\caption{Reduced $\chi^{2}$ value as a function of the \ce{CH4} mixing ratio for spectra within the MKO H band (in blue), MKO K band (in red), and our L band spectra (black). The L band curve has a more prominent local minimum that is not clearly seen at either H or K band, allowing for the \ce{CH4} abundances to be constrained on VHS 1256 b and PSO 318.5. The models used for this analysis are the best fit model with different \ce{CH4} mixing ratios.}
\label{fig:mixing ratio}
\end{figure}

If strong vertical mixing is present on numerous young, gas giant exoplanets, L band spectroscopy will be able to measure methane abundances for a wider range of objects at L to T transition temperatures than near-infrared spectroscopy alone. The best fit model from Section \ref{sec:models} was calculated with different \ce{CH4} mixing ratios above the quenching point. For each of those models, we calculated three separate reduced  $\chi^{2}$ values only considering spectra within the MKO H band, MKO K band, and the extent of our L band data. In Figure~\ref{fig:mixing ratio}, the reduced $\chi^{2}$ value versus \ce{CH4} mixing ratio is plotted for the H, K, and L band data of VHS 1256 b and PSO 318.5. The L band produces a more prominent local minimum along the methane mixing ratio curve than the H and K bands, showing that methane was detected and the abundance in VHS 1256 b and PSO 318.5 can be constrained.

\begin{figure}[ht]
\centering
\includegraphics[width = 3.1in]{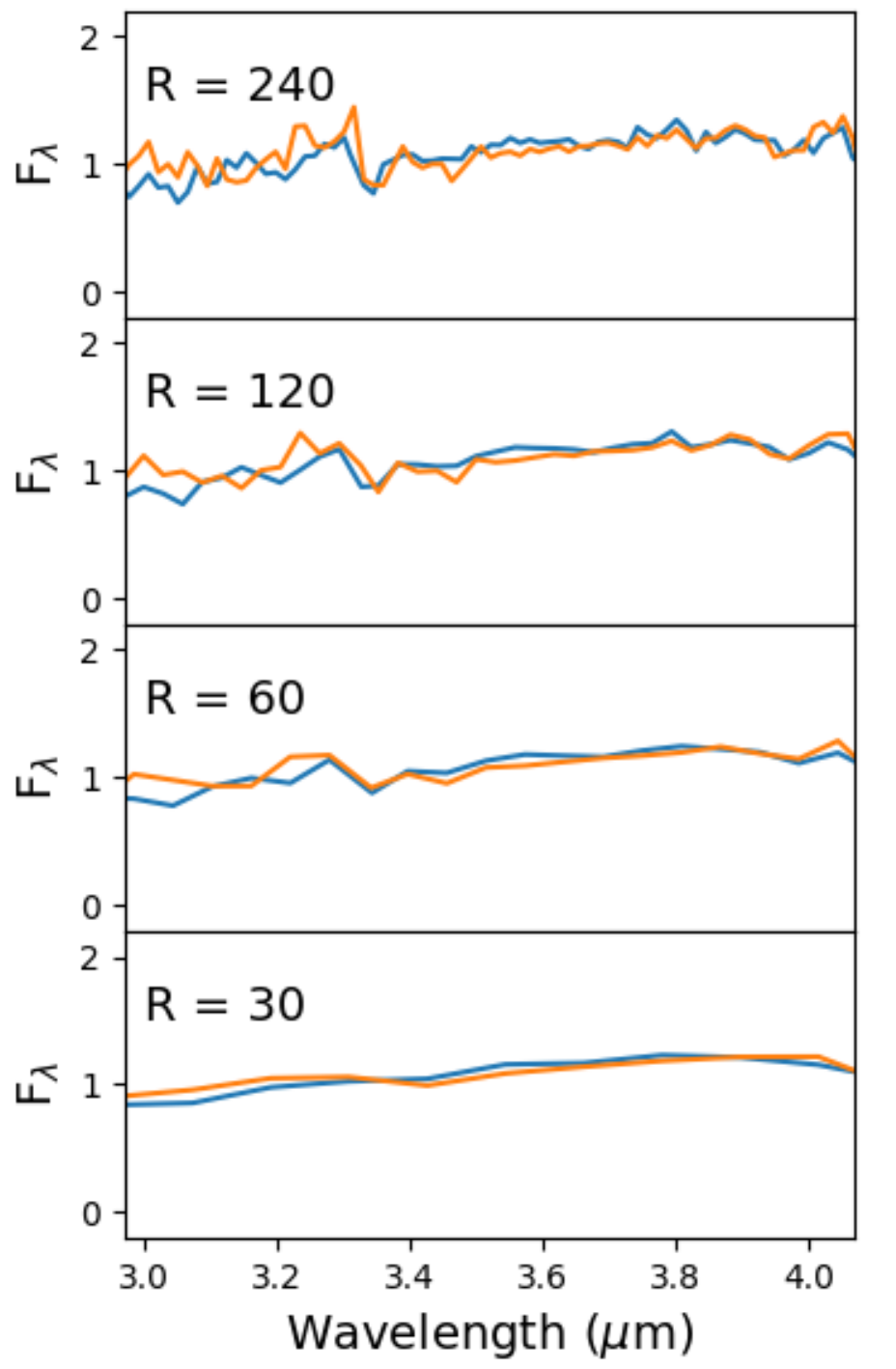}
\caption{The L band spectra of PSO 318.5 (blue) and VHS 1256 b (orange) binned to different resolutions. The spectra are smoothed down with a boxcar kernel. Depending on the signal to noise of the detection, the Q-branch methane feature is hard to distinguish from the continuum below a resolution of $\sim$ 200. }
\label{fig:resolution}
\end{figure}

The methane feature at $\sim$3.3 $\mu$m eventually blends in with the continuum at very low abundances when observed at lower resolutions (Figure \ref{fig:resolution}). Future L band observations that will study methane at similarly low abundances need to be done at resolutions of $\sim$200 or higher in order to detect methane. Medium to high resolution slit spectroscopy of isolated or wide companion brown dwarfs can be done from the ground at L band with a number of instruments on medium to large sized telescopes \citep[e.g. NIRSPEC, SpeX, iSHELL, CRIRES, and GNIRS][]{1998SPIE.3354..566M, 2003PASP..115..362R, 2016SPIE.9908E..84R, 2004SPIE.5492.1218K, 2006SPIE.6269E..4CE}). Present-day integral field units that are focused on finding and characterizing exoplanets in the glare of their host stars mostly operate blueward of L band \citep[e.g. SPHERE, GPI, CHARIS][]{2008SPIE.7014E..18B,2014PNAS..11112661M,2017SPIE10400E..16G}. Currently there is one exoplanet imaging IFU \citep[ALES][]{2015SPIE.9605E..1DS} capable of low-resolution L band spectroscopy. Higher spectral resolutions at L band are possible with lenslet and lenset/slicer integral field units \citep{2018arXiv180803304S}. Fiber injection units can be placed behind high contrast imaging systems to obtain even higher spectral resolutions of exoplanets \citep{2016SPIE.9911E..2TW, 2017SPIE10400E..29M}.

\subsection{Future JWST Observations}
\label{sec:jwst}

The \textit{James Webb Space Telescope} will be able to obtain high-fidelity spectra of brown dwarfs across a broad wavelength range, including L band. JWST can constrain the abundances of important secondary gases (e.g. \ce{H2O}, \ce{CO}, \ce{CH4}, \ce{NH3}) for the entire temperature range of known brown dwarfs \citep{2002Icar..155..393L}. Medium resolution spectra over a broad wavelength range will be critical for adequately defining and measuring the C/O ratios and metallicities for brown dwarfs and exoplanets. There is evidence for silicate condensates in L to T transition objects from \textit{Spitzer} observations \citep{2006ApJ...648..614C} and it may be possible to determine the phase and composition of these silicates with Mid-Infrared Instrument (MIRI) spectra. Direct studies of condensates are important for testing our assumptions about clouds and the extent to which other physical effects (ex. temperature instabilities, cloud microphysics) significantly influence the atmospheres of brown dwarfs and exoplanets. \citep{2017ApJ...850...46T,2018ApJ...853L..30L,2018ApJ...854..172C}.

Combining data from different epochs (as done in this work) may not accurately represent the average state of an atmosphere and introduces flux calibration errors. Most brown dwarfs with time series photometric measurements appear variable at the few percent level and rotate on the order of hours \citep{2017AstRv..13....1B}. PSO 318.5 has a measured variability of 3 - 4 \% percent in several bands across the 1 $\mu$m - 5 $\mu$m range and these variations are suggested to be due to inhomogeneous cloud layers \citep{2015ApJ...813L..23B,2018AJ....155...95B}. No time series observations have been taken of VHS 1256 b. The MKO and 2MASS J, H, and K /K$_{s}$ near infrared photometry for PSO 318.5 are different by $\sim$35, $\sim$9, and $\sim$23 percent respectively. The same is also true for the VISTA and 2MASS J, H, and K photometry of VHS 1256 b, which are different by $\sim$37, $\sim$13, and $\sim$3 percent respectively. The photometry of both objects are not completely inconsistent due to low signal to noise, however they can lead to different colors and flux calibrations. This ultimately influences our interpretation of the atmospheric and evolutionary properties of both objects. JWST's Near-Infrared Spectrograph (NIRSPEC) sensitivity enables contemporaneous coverage over the 1 $\mu$m - 5 $\mu$m wavelength range, making spectroscopic monitoring of brown dwarfs feasible. Contemporaneous and precise, medium resolution data of VHS 1256 b and PSO 318.5 will provide better constraints on the average effective temperature and other atmospheric properties of both objects.

\subsubsection{High Contrast Imaging of Exoplanets and Exoplanetary Early Release Science Program}

The High Contrast Imaging of Exoplanets and Exoplanetary Systems with JWST Early Release Science (ERS) Program will utilize all of JWST's instruments to provide a useful data set for the astonomical community to assess JWST's ability to execute observations of exoplanetary systems \citep{2017jwst.prop.1386H}. VHS 1256 b is the first choice target for NIRSPEC and MIRI integral field unit (IFU) mode observations that will provide the first medium resolution (R $>$ 1,000), 0.7 $\mu$m - 28.0 $\mu$m spectrum of an exoplanet analog. 

The ERS spectrum will detect the majority of VHS 1256 b's luminosity, important carbon, oxygen, and nitrogen bearing gases, and silicate condensates if present. The resolution of the ERS spectrum is sufficient for detecting the methane features presented in this work and other minor/low abundance molecular absorbers. The spectrum will be able to constrain all of the important carbon and oxygen bearing gases required to make an estimate for the C/O ratio of VHS 1256 b.

For gaseous Solar System planets, the $K_{zz}$ profile of an object may change relatively quickly over a scale height or yield different results depending on which non-equilibrium gases are measured \citep{1998Icar..132..176S, 2002Icar..159..112L}. NIRSPEC and MIRI measurements of \ce{CO} and \ce{CH4} at multiple wavelengths can test the strength and vertical extent of $K_{zz}$ along an assumed pressure vs. temperature profile for VHS 1256 b. The ERS spectrum will be an enormous leap in the quality of data we have to study a directly-imaged exoplanet analog, and will allow us to test, compare, and improve the array of models and retrievals used to characterize the atmospheres of brown dwarfs and exoplanets.

\section{Summary and Conclusions}
\label{sec:summary}
Low gravity objects like 2MASS 1207 b and the HR 8799 planets appear redder on the color magnitude diagram relative to field brown dwarfs. Their near infrared spectra lack strong signatures of methane hinting that these atmospheres are out of chemical equilibrium. This non-equilibrium chemistry can be explained by vertical mixing exchanging warm carbon monoxide-rich gas with cool methane-rich gas. The opacity of methane is greatest within the L band and dominates over other secondary gases such as carbon monoxide and water, making it an excellent wavelength region for constraining methane in low gravity L to T transition objects.

In this work, we detect methane in R $\sim$1,300 L band spectra of the planetary mass companion VHS 1256 b, and the young, free floating planetary mass object PSO 318.5, which share the same color space as the HR 8799 planets. The L band spectra of these two objects provide an early look into the composition and cloud properties of self luminous, gas giant exoplanets that we may characterize with large ground based telescopes in the future.

The results of this paper are:
\begin{itemize}

\item We detect low abundances of \ce{CH4} in L band spectra of VHS 1256 b and PSO 318.5. Both objects have spectral slopes that are relatively flat, similar to the 3 $\mu$m - 4 $\mu$m photometry of HR 8799 c and d featured in \cite{2014ApJ...792...17S}. HR 8799 c and d have deeper troughs through 3.3 $\mu$m, suggesting that these planets have more methane and/or less vertical mixing than VHS 1256 b and PSO 318.5.

\item The reduced quantities of \ce{CH4} detected in VHS 1256 b and PSO 318.5 suggest non-equilibrium chemistry between \ce{CO} and \ce{CH4}. The upper atmospheres of the best fit models depart from equilibrium abundances of \ce{CH4} by factors of 10 - 100.  An eddy diffusion coefficient (K$_{zz}$) of $10^{8}$ cm$^{2}$ s$^{-1}$ is required to match the depths of the 3.3 $\mu$m \ce{CH4} features.

\item The best fit model to VHS 1256 b and PSO 318.5 is a low gravity, 1240K object, with photospheric clouds. This model produces a reasonable radius for PSO 318.5 that is consistent with evolution models \citep{1998A&A...337..403B}. VHS 1256 b's spectrum looks remarkably similar to PSO 318.5, however, its uncertain distance manifests as an unconstrained radius. Future distance and age measurements will be able to connect these objects and others from an evolutionary standpoint.

\item Medium resolution, L band spectroscopy can detect low abundances of \ce{CH4} at 3.3 $\mu$m even when the \ce{CH4} abundances are too low to be detectable at 1.6 $\mu$m in medium resolution H band spectra. For objects similar to VHS 1256 b and PSO 318.5, a R $\sim$ 200 is necessary to detect \ce{CH4} at L band.

\end{itemize}

All known directly imaged exoplanets fall within the temperature range of brown dwarfs. A handful of brown dwarfs even share the same infrared colors as exoplanets and they are incredibly useful laboratories for studying the variety of atmospheric abundances and processes that are likely to be seen on gas giant exoplanets.

\section{Acknowledgements}
The authors wish to recognize and acknowledge the very significant cultural role and reverence that the summit of Maunakea has always had within the indigenous Hawaiian community.  We are conducting observations from this mountain, which is colonized land. This work benefited from the Exoplanet Summer Program in the Other Worlds Laboratory (OWL) at the University of California, Santa Cruz, a program funded by the Heising-Simons Foundation. This work was funded by the NSF Planetary Astronomy, Awards 1614320 and 1614492.  J.M.S. is supported by NASA through Hubble Fellowship grant HST-HF2-51398.001-A awarded by the Space Telescope Science Institute, which is operated by the Association of Universities for Research in Astronomy, Inc., for NASA, under contract NAS5-26555. Barman is supported by NSF AAG awards 1405505 and 1614492 and NASA XRP award NNX17AB63G

Software  The following software packages and programming languages made this research possible: Astropy, a community-developed core Python package for Astronomy \citep{2018arXiv180102634T}, \texttt{Matplotlib} \citep{4160265}, \texttt{SciPy} \citep{Scipy}, \texttt{Numpy} \citep{2011arXiv1102.1523V}, and \texttt{Python 2.7.10}

B.E.M. would like to thank: Gregg Doppman for answering questions about NIRSPEC. Bartek Gauza who re-reduced the VHS 1256 b near infrared spectra for our analysis.  M. Buffaloe for reminding me to be my authentic self. Solange Knowles for creating \textit{A Seat at the Table}.

\bibliographystyle{apj}
\bibliography{bibliography.bib}

\end{document}